\documentclass{article}
\usepackage[utf8]{inputenc}
\usepackage{graphicx}
\usepackage{caption}
\usepackage{wrapfig}
\usepackage{soul}

\usepackage[T1]{fontenc}
\usepackage[utf8]{inputenc}
\usepackage{authblk}

\usepackage[letterpaper, margin=1in]{geometry}
\usepackage[notes,backend=biber]{biblatex-chicago}
\bibliography{ref}
\usepackage{multicol}

\usepackage{xcolor}

\title{Risks to Zero Trust in a Federated Mission Partner Environment}

\author[1,2]{Keith Strandell\thanks{kes828@msstate.edu}}
\author[2]{Sudip Mittal\thanks{mittal@cse.msstate.edu}}

\affil[1]{United States Air Force}
\affil[2]{Computer Science \& Engineering, Mississippi State University}

\date{}

\begin{document}

\maketitle

\section*{Abstract}
Recent cybersecurity events have prompted the federal government to begin investigating strategies to transition to Zero Trust Architectures (ZTA) for federal information systems. Within federated mission networks, ZTA provides measures to minimize the potential for unauthorized release and disclosure of information outside bilateral and multilateral agreements. When federating with mission partners, there are potential risks that may undermine the benefits of Zero Trust. This paper explores risks associated with integrating multiple identity models and proposes two potential avenues to investigate in order to mitigate these risks.

\section{Introduction \& Background}
Within days following the cyberattack on the Colonial Pipeline, Joseph R. Biden Jr., President of the United States, signed into effect Executive Order 14028: Improving the Nation’s Cybersecurity \autocite{Biden}. Prompted by recent ``sophisticated and malicious'' cyberattacks, the order acts as a catalyst for federal agencies to take necessary and immediate steps to coordinate with industry on improving information sharing, adopting best practices and, among other things, migrating Federal Information Systems from perimeter-based security to a Zero Trust Architecture (ZTA). The foundational elements of Zero Trust are micro-segmentation and a well-informed trust algorithm. When effectively implemented with data tagging, Zero Trust provides a strong compartmentalization model that lends itself to federated mission partner environments. However, given an environment where the mission partner is responsible for bringing to the table their own identity model, consideration must be given to risks associated with federating multiple mission partners.

In this paper, we investigate the risks associated with a multi-partner environment built on ZTA that federates with each mission partner's identity model. For the purposes of isolating the impact of federated identities, the operating assumption is the environment has fully implemented micro-segmentation and data tagging such that the primary risks are associated with the integration of multiple identity models. In addition to assessing the risks, the paper recommends two potential areas of investigation that may alleviate some of the risks associated with this architecture.  

\section{Mission Partners and Data Protection}
The international mission partners Combatant Commands (COCOMs) work with take on a variety of forms, the most obvious being foreign military. However, there is a significant amount of cooperation that occurs with other agencies. In January ‘10, U.S. Southern Command (USSOUTHCOM) responded to a request for earthquake relief support in Haiti. This Humanitarian Assistance and Disaster Response (HA/DR) operation required coordination with multiple international organizations, including foreign government agencies, nongovernment agencies and foreign militaries. In order to effectively share information, data was kept unclassified to the maximum extent possible and public platforms were used for dissemination \autocite{arroyoCenter}. Another example of cooperation with international partners can be found in a recent partnering between U.S. Africa Command (USAFRICOM), the International Criminal Police Organization (INTERPOL) and local law enforcement from several West African nations. The operation targeted illegal fishing and “other maritime crimes” along the West African coast \autocite{maritime}. Not only do these mission sets require sharing of unclassified data, but they also demonstrate the potential for both persistent and transient user bases operating in the same environment.

Attempting to create a collaborative environment to facilitate data sharing that allows for multiple missions and user bases exacerbates the need for effective controls to prevent the unauthorized release and disclosure of information such as Controlled Unclassified Information (CUI). For example, data controlled as NOFORN is not releasable to foreign mission partners; however, it may need to reside in this environment due to a need to release to non-foreign entities such as the Federal Emergency Management Agency. Similarly, data controlled as “CUI//REL TO USA, FVEY” is releasable to members of the Five Eyes alliance \autocite{CUIppt}. A comparable protection requirement exists for mission partner data. The Mission Partner Environment framework is designed to facilitate collaboration and sharing with “participants within a specific partnership or coalition” \autocite{CJCSI629001}. The implication is there is a requirement to ensure data is only shared within designated groups. For example, assume there are existing agreements between the United States, country A and country B, as depicted in Figure 1. In this image, the overlapping areas represent shared data based on these partnerships. Each country contributing data to the environment expects the information it uploads to the system to be protected accordingly. That is to say, data transferred to the United States as part of a bilateral agreement with country A must not be released to country B without the express consent of country A.\\
\begin{figure}[h!]
    \includegraphics[width=0.5\linewidth]{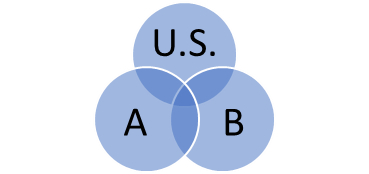}
    \centering
    \caption{Multi-country data sharing partnerships.}
    \label{fig:sharing}
\end{figure}

\section{Zero Trust and Federation}
The operating assumption in ZTA is that the network is compromised and therefore steps must be taken to minimize the potential impact of unauthorized access. Through micro-segmentation and data tagging, a ZTA can provide a framework where compartmentalization is baked into the security model. The result is smaller trust zones which reduce the potential for lateral movement of an adversary exploiting a vulnerability (see Figure 2). However, to fully realize the benefits, the ZTA must also implement a trust algorithm that takes in relevant data feeds to provide continuous authentication and authorization decisions on access requests. A robust trust algorithm will have access to contextual information on the requesting entity and device, the target resource, resource access policies and threat intelligence \autocite{NIST}. Access to these information feeds provide a more complete view of the request and associated risks. For example, consider the ability to access data related to the requesting entity's device configuration in order to compare with data on known configurations in order to predict the level of vulnerability associated with the device \autocite{patent}. Such an assessment increase the insight into the risk of a given request.\\
\begin{figure}[h!]
    \includegraphics[width=0.5\linewidth]{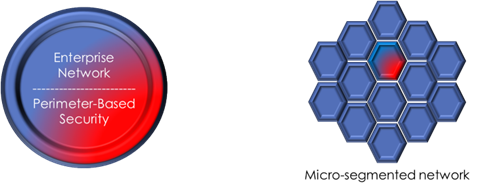}
    \centering
    \caption{Lateral Movement in Perimeter Network vs. Zero Trust.}
    \label{fig:Lateral Movement}
\end{figure}
Federation in ZTA poses an interesting conundrum because in an architecture that strives to remove trust, it introduces an inherent trust between the federated organizations. Here, the mission partners act as identity providers and are responsible for authenticating their users. Once authenticated, the identity is securely transferred to support the trust algorithm making the authorization decision. This model eliminates the need for the user to maintain security information related to a separate identity which can reduce the risk of compromise associated with user behavior; however, it can undermine the rigor of the trust algorithm by preventing access to contextual information related to the requesting entity.

\section{Risks Federating with Mission Partner Identity Solutions}
Zero Trust touts a robust trust algorithm rooted in the ability to verify a user’s identity; however, in a federated model, the algorithm is only as strong as the weakest identity solution. Ideally this risk would be mitigated by defining a establishing a minimum baseline that all partners must adhere to; however, when balancing risk and mission requirements, mission may take priority and drive risky behavior or decisions. As such, there is the potential for integration with substandard models. In this scenario, the mission partner’s identity potentially undermines the trust algorithm and puts at risk any data shared with them. Aside from the potential of a substandard model, there is also a lack of context that comes with federation. A robust trust algorithm requires a significant amount of user and device information to make informed decisions; however, the ability to access the contextual information around the requesting entity is a challenge in federated models \autocite{context}.  

The effective establishment and enforcement of a common baseline increases the security of the system; however, it does not address vulnerabilities in the supply chain.  On December 8, 2020, it was discovered that SolarWinds had been compromised. The hack, attributed to Russia, affected approximately 17,000 of SolarWinds clients, including several federal agencies such as the Department of Homeland Security and the Department of Defense. The attackers targeted a third-party vendor, Orion, that had a long-standing relationship with SolarWinds. By infiltrating Orion, when clients of SolarWinds updated their software, they inadvertently loaded malware onto the devices giving hackers access to their networks, which in many instances resulted in significant data breaches \autocite{SolarWinds}. The attack is significant in that it focused on popular network infrastructure devices which allowed for the vast attack surface. A comparable attack on identity solutions for participating organizations in a federated environment could introduce an attack vector to the environment and expose data shared with the mission partner. This risk is amplified by strategic competitors’ ability to leverage the Diplomatic, Informational, Military and Economic (DIME) framework to deliberately position state sponsored technology that provides them covert access.

Strategic competitors, such as the People’s Republic of China (PRC) and Russia are actively exercising DIME strategies to advance their influence in regions around the world. General Towhsend noted in his statement to the House Armed Services Committee that both countries have an ``inside track'' in central and southern Africa. He also states that Russia is actively buying influence in the region and the PRC is investing billions in infrastructure and development in Africa \autocite{PostureAFRICOM}. Within the USSOUTHCOM’s AOR, the PRC holds \$165B in loans and is using COVID-19 as a pretext to further indebt the region while enhancing their integration with the infrastructure and technology. For example, as part of their COVID-19 response, the PRC was offering to donate Huawei technology \autocite{PostureSOUTHCOM}. The significant investments these competitors are infusing into the region provide the pretense to gain access to senior government officials with the leverage to secure deals that further embed their technology or allow insight/access to processes such as identity management. The infrastructure investments in these regions serve, at a minimum, as a method to increase reliance and influence; however, they also introduce the potential supply risk noted above. Specific to the PRC, there are concerns related to the Military-Civil Fusion Strategy and how involved vendors such as Huawei are with the People’s Liberation Army and the extent to their collaborations \autocite{Myths}. This concern is furthered by incidents that suggest not only security issues but the intentional inclusion of surveillance capabilities that lend themselves to espionage\autocite{Kania}. While the Huawei push is focused on 5G, the concern extends to any presumably state-sponsored technology that may serve as critical infrastructure to mission partner networks and increase the risks to the federated model.

\section{Conclusion}
The transition from perimeter-based cybersecurity to ZTAs should result in significant improvements in the overall security posture of enterprise networks. Specifically, it shows promise in the realm of multi-national operations where cooperation can often be borne out of necessity and built on a tentative trust between mission partners. The inherent compartmentalization of a robust ZTA lends itself well to an environment rooted mission partners’ trust that their data is protected from unauthorized release and disclosure. Unfortunately, the benefits of Zero Trust can be undermined by the federating of multiple identity models. Strategic competitors are actively employing the DIME framework to enhance their regional footprints and advance their influence and deploy state-sponsored technologies. These activities increase the opportunities for social engineering, political influence, and clandestine cyber operations. Some of the risks can be mitigated by limiting federation to mission partners with known, trusted architectures and limited ties to strategic competitors while offering to host all other partners. This model, however, has the potential to be compromised when mission requirements outweigh the cybersecurity risks. To further secure the environment’s security posture, additional measures should be investigated.

Two promising options for enhancing the authentication model that could be investigated as augmenting technologies are blockchain and adaptive neuro-fuzzy inference systems. Blockchain gained popularity as the digital ledger supporting bitcoin transactions. However, recent efforts using blockchain for authentication as part of a self-sovereign identity model have been demonstrated. In 2019, a group of credit unions piloted the use of blockchain and noted the improvement in the authentication model could reduce a credit union’s annual fraud expenses by \$150K just by reducing the authentication risks tied to call centers\autocite{CUId}. ANFIS is a machine learning model that integrates adaptive neural networks with a fuzzy inference system. In a study on the potential use to support “continuous implicit authentication” on mobile devices, ANFIS was used to learn user behaviors to support implicit user authentication and identification of both informed and uninformed adversary attacks. While the model showed a 5\% increase in user recognition, the improvement in informed adversary attacks was negligible and it underperformed on identifying uninformed adversary attacks \autocite{ANFIS}. The ANFIS architecture does show promise for user authentication on mobile devices; however, if paired with an identity model, it may be used as part of an enterprise authentication solution that focuses on learning archetype behaviors in order to identify when a user’s behavior deviates from the normal behaviors of users assigned to the same role. \\\\

\noindent\textbf{DISCLAIMER}: {The views expressed in this work are those of the author(s) and do not reflect the official policy or position of the United States Military Academy, the Department of the Army, the Department of the Air Force or the Department of Defense.}

\section*{Biographies}
\begin{wrapfigure}{1}{0.25\textwidth}
\captionsetup{labelformat=empty,listformat=empty}
    \includegraphics[width=0.9\linewidth]{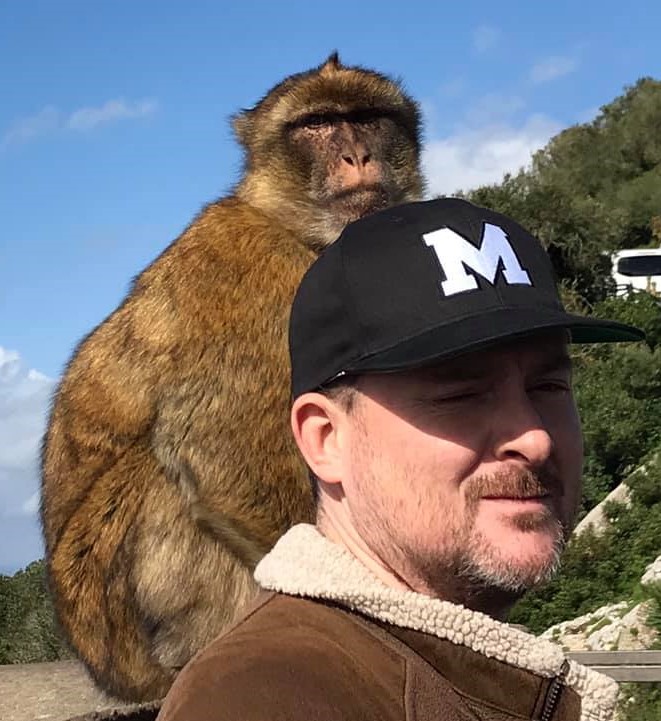}
    \caption*{Keith Strandell}
    \label{fig:wrapfig}
\end{wrapfigure}
\textbf{Keith Strandell} is the Materiel Leader for the Royal Saudi Air Force AWACS Modernization Program and previously served as the Materiel Leader for life cycle management of enterprise applications within the Enterprise Services portfolio within the United States Air Force's Enterprise Information Technology (EIT) construct, including capabilities such as the Air Force’s Office 365 instance in the Impact Level 5 environment. Keith started his career as a Communication and Information Officer in the Air Force leading the Network Security and Information Assurance offices at Travis Air Force Base.  From there, he transitioned to a Technical Program Management role leading various hardware and software acquisition/development efforts related to battle control, intelligence and EIT systems. Included in this time frame were 4 years managing Foreign Military Sales cases across the Pacific, European and Central Commands. Keith is currently pursuing a Ph.D. in Computer Science \& Engineering at Mississippi State University.\\\\

\begin{wrapfigure}{1}{0.25\textwidth}
\captionsetup{labelformat=empty,listformat=empty}
    \includegraphics[width=0.9\linewidth]{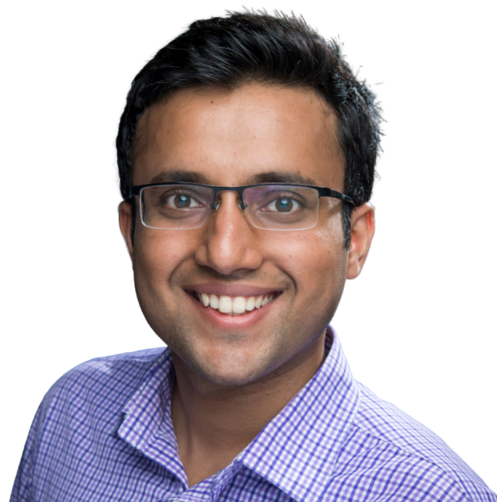}
    \caption*{Sudip Mittal}
    \label{fig:wrapfig}
\end{wrapfigure}
\noindent\textbf{Sudip Mittal} is an Assistant Professor in the Department of Computer Science \& Engineering at  Mississippi State University. He graduated with a Ph.D. in Computer Science from the University of Maryland Baltimore County in 2019. His primary research interests are cybersecurity and artificial intelligence. Mittal’s goal is to develop the next generation of cyber defense systems that help protect various organizations and people. At Mississippi State, he leads the Secure and Trustworthy Cyberspace (SECRETS) Lab and has published over 70 journals and conference papers in leading cybersecurity and AI venues. Mittal has received funding from the NSF, USAF, USACE, and other Department of Defense programs. He also serves as a Program Committee member or Program Chair of leading AI and cybersecurity conferences and workshops. Mittal’s work has been cited in the LA times, Business Insider, WIRED, the Cyberwire, and other venues. He is a member of the ACM and IEEE.

\printbibliography
\end{document}